\documentclass[twocolumn,showpacs,preprintnumbers,amsmath,amssymb,aps,prb]{revtex4}
\usepackage{graphicx}
\begin{document}

\title{
Reversible Ratchet Effects for Vortices in Conformal Pinning Arrays 
} 
\author{
C. Reichhardt, D. Ray,
and C. J. Olson Reichhardt
} 
\affiliation{
Theoretical Division,
Los Alamos National Laboratory, Los Alamos, New Mexico 87545 USA
} 

\date{\today}
\begin{abstract}
A conformal transformation of a uniform triangular pinning array
produces a structure called a conformal crystal which preserves
the six-fold ordering of the original lattice 
but contains a gradient in the pinning density.
Here we use numerical simulations to 
show that vortices in type-II superconductors driven with an
ac drive over
gradient pinning arrays produce the most pronounced ratchet
effect over a wide range of parameters for a conformal array,
while square gradient or random gradient arrays with equivalent
pinning densities give reduced ratchet effects.
In the conformal array,
the larger spacing of the pinning sites in the direction transverse to
the ac drive permits easy funneling of interstitial
vortices for one driving direction, producing the enhanced ratchet
effect.
In the square array, the transverse spacing between pinning sites
is uniform, giving no asymmetry in the funneling of the vortices as
the driving direction switches, while in the random array, there are
numerous easy-flow channels present for either direction of drive.
We find multiple ratchet reversals in the conformal arrays
as a function of vortex density and ac amplitude,
and correlate the features with a reversal in the vortex ordering, which
is greater for
motion in the ratchet direction.
The enhanced conformal pinning
ratchet effect can also be realized for colloidal particles moving over a
conformal
array, indicating the general usefulness of conformal structures for controlling the motion of particles.  
\end{abstract}
\maketitle

\section{Introduction}
When an assembly of particles are placed in an
asymmetric potential, a net dc particle flow can arise due to a
ratchet effect that occurs
when an ac drive is applied or when the substrate
is periodically switched on and off in the presence
of a thermal bath \cite{1,2}.    
Ratchet effects on asymmetric substrates have been extensively studied 
in colloidal systems \cite{3,4,5},
granular matter \cite{6,7}, and polymers \cite{8,9}. 
Ratchet effects also appear in ac-driven vortices in type-II superconductors
in the presence of an  
asymmetric substrate \cite{6a,7a,8a,9a,10,11},
such as a quasi-one-dimensional periodic array produced by asymmetrically
modulating the sample thickness \cite{6a,12,13,14,15},
etching funnel-shaped channels for vortex flow \cite{7a,16,17,18,19,20},
introducing asymmetry to the 
sample edges \cite{21}, or   
adding periodic pinning arrays in which the individual pinning sites have 
some form of intrinsic asymmetry \cite{9a,10,22,23,24,25,26,27,28,29,30}. 
At lower vortex densities when collective interactions between vortices
are weak, the ratchet effect produces a dc flow of vortices in the
easy flow direction of the asymmetric substrate; however, when collective
effects are present it is possible to
have reversals of the ratchet effect 
where for one set of parameters the vortices move in the easy direction while
for another set of parameters they move in the
hard direction 
\cite{9a,10,11,12,14,22,23,24,25,26,27,28}.
A ratchet effect can also be produced by a pinning array containing
symmetric pinning sites arranged with a density gradient.
Olson {\it et al.} \cite{8a} first studied vortex ratchet effects for
random gradient array pinning geometries and found that the vortices
undergo a net dc flow in the easy direction.  Experiments and
simulations later showed that for a square array of pinning sites with
constant pinning density but with a gradient in pinning site size,
a variety of forward and reverse
vortex ratchet behaviors occur \cite{31}. 
Experiments on triangular pinning arrays with a density gradient
also revealed a forward ratchet effect at low fields with a reversal 
at higher fields \cite{32,33}.

\begin{figure}
\includegraphics[width=3.5in]{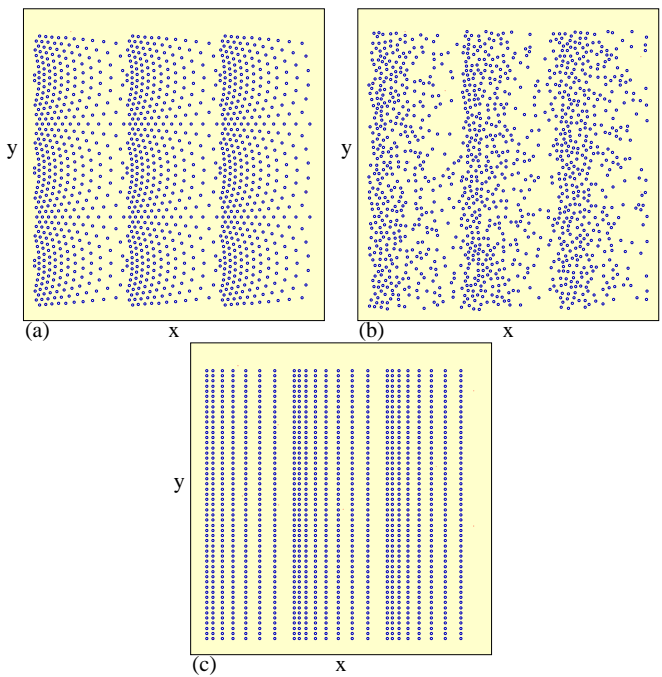}
\caption{ 
The system geometry showing the 
locations of the pinning sites for samples with $n_{p} = 1.0$.
Pinning gradients run along the $x$ direction.
(a) Conformal pinning (Conf). 
(b) Random pinning with a periodic gradient (RandG).
(c) Square pinning with a gradient (SquareG). 
For each sample we apply an ac drive along the $x$ direction and
measure the average net vortex displacements
$\langle \Delta X\rangle$ in the $x$ direction.
}
\label{fig:1}
\end{figure}

Recently a new type of pinning geometry was proposed that is constructed
by conformally transforming a triangular pinning lattice to create
what is called a conformal pinning array,
abbreviated Conf in this work \cite{34,35,36}. 
As in the original triangular lattice, each pinning site
in the transformed array 
has six neighbors separated by 60$^\circ$; however, the distance to each
neighbor is
no longer constant, producing a density gradient in the
pinning sites \cite{37,38}.
In Fig.~\ref{fig:1}(a) we show a periodic lattice composed of
three conformal pinning arrays with the same orientation.
Experimental structures with nearly conformal geometries have been
observed
for magnetically interacting particles subjected to
a gravitational force,
and due to the arching nature of the conformal array,
the magnetic conformal crystals were dubbed
gravity's rainbow structures \cite{38}. 
Similar conformal structures have also been 
studied in  foams \cite{40,41} and 
in charged particle
ordering in confined geometries \cite{41}. 
In the superconducting system,  
an arrangement of two Conf arrays placed with their minimum
pinning density regions
in the center of the sample produces an enhanced critical current or
depinning force compared to an equivalent number of pinning sites
placed in a uniform periodic, uniform random, or random density
gradient array \cite{34}.
The enhancement results both from the natural density gradient formed
by the vortices as they enter the sample from the edges and form a Bean
state, and from the preservation of local six-fold ordering in
the Conf array \cite{34,36}.
The suppression of 
easy vortex flow channels which arise in 
periodic and random arrays
by the arching conformal structure
also plays a role in the enhancement. 
Only at the integer matching fields do the
periodic pinning arrays produce higher critical currents than the Conf array
\cite{36}.
Subsequent experimental studies confirmed that the 
Conf array produces enhanced pinning over a wide range of
fields compared to uniform periodic pinning or uniform random 
pinning arrangements \cite{42,43}. 
Other experiments have shown enhanced critical
currents in systems with periodic arrays when a 
gradient in the pinning density is introduced \cite{44}. 
There have also been studies of
hyperbolic-tesselation arrays 
which have a gradient in the pinning density \cite{45}. 

Since conformal pinning arrays have an 
intrinsic asymmetry, it is natural to ask whether a ratchet effect can
occur under application of an ac drive, and if so,
whether this ratchet effect would be
enhanced compared to other 
pinning array geometries with density gradients, 
or whether ratchet reversals could be possible. 
In Fig.~\ref{fig:1} we illustrate
three examples of the gradient pinning array geometries
we consider in this work:
conformal pinning (Conf) [Fig.~\ref{fig:1}(a)], 
random pinning with a gradient (RandG) [Fig.~\ref{fig:1}(b)],
and a square pinning array containing a gradient along
the $x$ direction (SquareG) [Fig.~\ref{fig:1}(c)].
In each case, both the gradient and the applied ac drive are along
the $x$ direction, while
the easy-flow direction for vortex motion is in the
negative $x$ direction.
In addition to superconducting vortex realizations of these geometries,
similar pinning arrangements could also be created in colloidal
systems using optical trap arrays.

\begin{figure}
\includegraphics[width=3.5in]{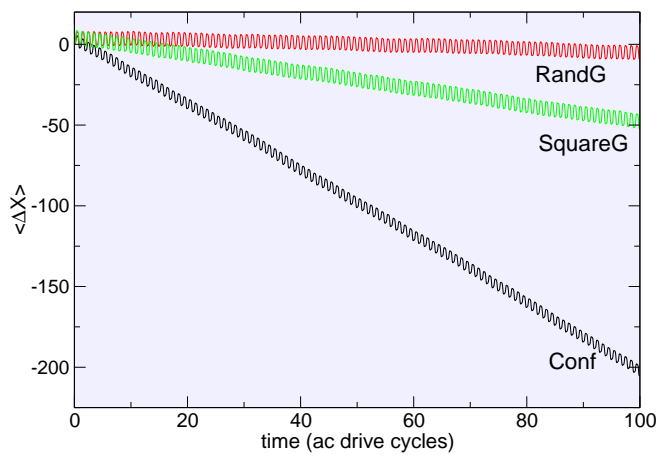}
\caption{The average net displacement $\langle \Delta X\rangle$ vs time measured in
ac drive cycles
for the systems in Fig.~1 at $B/B_{\phi} = 1.0$, $F_{p} = 1.0$, 
and $F_{ac} = 0.55.$
Bottom black curve: Conf array; middle green curve: SquareG array; upper red curve:
RandG array.
Here the Conf
array produces a ratchet that is
four times more effective than the SquareG array
and 20 times more effective than the RandG array. 
}
\label{fig:2}
\end{figure}

\section{Simulation and System}   
We consider a two-dimensional system with periodic boundary 
conditions in the $x$ and $y$ directions.
The sample size 
is $L \times L$ with $L=36\lambda$, where
distance is measured in units
of the London penetration depth $\lambda$. 
The applied magnetic field 
is perpendicular to the system in the ${\bf \hat z}$ direction.  
Our results apply to the London limit regime in which 
the vortices can be treated as rigid objects,
when the coherence length $\xi$ is much smaller than $\lambda$. 
The pinning sites are modeled 
as in previous studies of conformal pinning arrays \cite{34,35,36} 
by non-overlapping parabolic circular traps 
with radius $R_{p}$ and a maximum pinning 
force of $F_{p}$.  
We place the $N_p$ pinning sites in a conformal array (Conf), as described in
previous work \cite{34},
in a random arrangement with a gradient (RandG), or in a square array with a
density gradient along the $x$ direction (SquareG).
The width of each pinning
array segment is $a_{p} = 12\lambda$, and the segments are
repeated three times across the sample as shown in Fig.~\ref{fig:1}. 
The total density of the pinning sites 
is $n_{p} =  N_{p}/L^{2} = 1.0$.
The sample contains 
$N_{v}$ vortices and we
measure the magnetic field in units of $B/B_{\phi}$, 
where $B_{\phi}$ is the matching field at which there is
one vortex per pinning site.      
We obtain
the initial vortex configuration by 
annealing from a high temperature molten state and cooling to 
$T=0$ or
to a low but finite fixed temperature. 
After annealing, we apply an ac driving force to all the vortices.

The dynamics of an individual vortex $i$ 
is obtained by integrating the following overdamped equation of
motion:
\begin{equation}  
\eta \frac{d {\bf R}_{i}}{dt} = 
 {\bf F}^{vv}_{i} + {\bf F}^{vp}_{i} + {\bf F}^{ac}_{i} + {\bf F}^{T}_{i} \ .
\end{equation} 
Here  $\eta$ is the damping constant
which is set equal to 1.  
The repulsive vortex-vortex 
interaction force is
given by 
${\bf F}^{vv}_{i} = \sum_{j\neq i}F_{0}K_{1}(R_{ij}/\lambda){\hat {\bf R}_{ij}}$,
where ${\bf R}_{i}$ is the location of vortex $i$,
$K_{1}$ is the modified Bessel function, 
$R_{ij} = |{\bf R}_{i} - {\bf R}_{j}|$,
$ {\hat {\bf R}_{ij}} = ({\bf R}_{i} - {\bf R}_{j})/R_{ij}$, 
$F_{0} = \phi^{2}_{0}/(2\pi\mu_{0}\lambda^3)$, 
 $\phi_{0}$ is the flux quantum, and $\mu_{0}$ is the permittivity.    
The vortex-pinning interaction force is
${\bf F}^{vp}_{i} = \sum^{N_{p}}_{k= 1}(F_{p}R^{(p)}_{ik}/r_{p})\Theta((r_{p}  
-R^{(p)}_{ik})/\lambda){\hat {\bf R}^{(p)}}_{ik}$,  
where $\Theta$ is the Heaviside step function, 
$r_{p} = 0.25\lambda$ is the pinning radius, $F_{p}$ is the pinning strength, 
${\bf R}_k^{(p)}$ is the location of pinning site $k$,
$R_{ik}^{(p)} = |{\bf R}_{i} - {\bf R}_{k}^{(p)}|$, and
$ {\hat {\bf R}_{ik}^{(p)}} = ({\bf R}_{i} - {\bf R}_{k}^{(p)})/R_{ik}^{(p)}$. 
All forces are measured in units of $F_{0}$ and lengths in units of $\lambda$. 
Thermal forces are represented by Langevin kicks ${\bf F}^{T}_{i}$ with the properties 
$\langle F^{T}_{i}(t)\rangle = 0$ and 
$\langle F^{T}_{i}(t)F^{T}_j(t^{\prime})\rangle = 2\eta k_{B}T\delta_{ij}\delta(t-t^{\prime})$,
where $k_B$ is the Boltzmann constant.   
The ac driving force is 
${\bf F}_{ac} = F_{ac}\sin(\omega t){\hat {\bf x}}$ where $F_{ac}$ is the ac amplitude. 
To characterize the ratchet effect we measure the average net displacement of
all vortices from their starting positions as a function of time,
$\langle \Delta X\rangle=N_v^{-1}\sum_{i=1}^{N_v} (x_i(t)-x_i(t_0))$, where
$x_i(t)$ is the $x$ position of vortex $i$ at
time $t$ and $t_0$ is an initial reference time.
This measure produces a sinusoidal signal, as shown in Fig.~\ref{fig:2};
the presence of a net drift indicates that a ratchet effect is
occurring.
We condense this information into a single number $X_{\rm net}$, the value of
$\langle \Delta X\rangle$ at $t-t_0=50$ ac drive cycles.
Except where otherwise noted,
we consider a fixed ac frequency of $\omega=0.04$ 
and a time step of $\delta t=0.02$, so that a single drive cycle
has a period of
8000 simulation time steps.

\section{Ratchet Effect}

In Fig.~\ref{fig:2} we plot the average net displacement
$\langle \Delta X\rangle$ versus time
for Conf, RandG, and SquareG arrays with $B/B_\phi=1.0$, $F_p=1.0$, and
$F_{ac}=0.55$ 
during 100 ac drive cycles.
Here the overall drift of each curve indicates that
all the arrays produce a ratchet effect with the vortices translating in the
negative $x$ direction.
The Conf array generates the largest ratchet effect, with the vortices translating distances
up to $2\lambda$  per drive cycle.
The ratchet effect for the Conf array is about four times larger
than that of the SquareG array  and
20 times larger than that of the RandG array.

\begin{figure}
\includegraphics[width=3.5in]{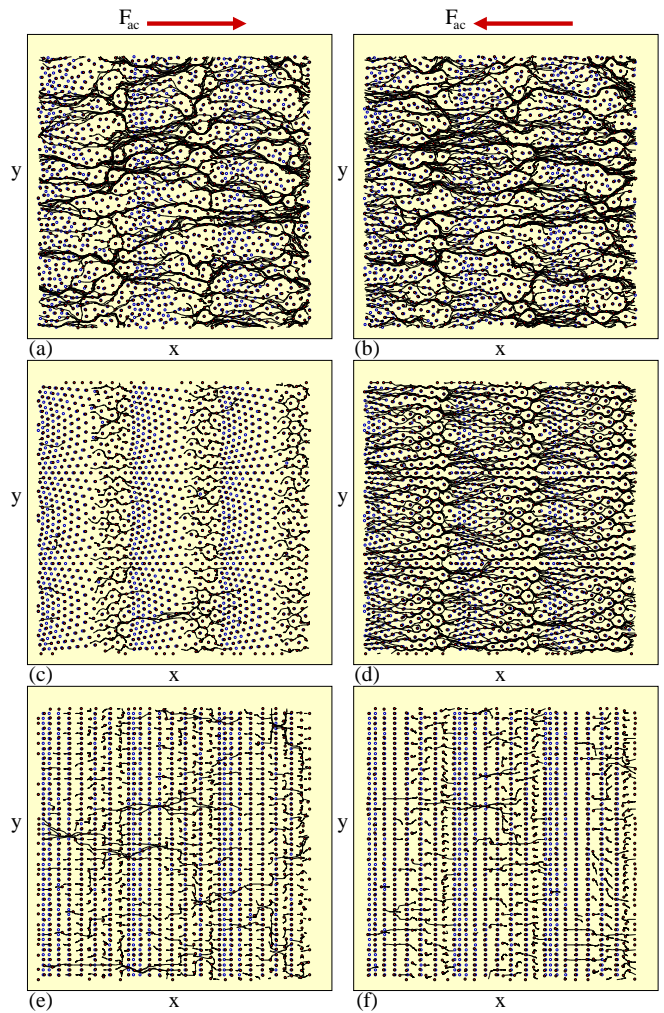}
\caption{ 
  The vortex locations (red dots), pinning site locations (open
  blue circles), and vortex trajectories (black lines)
  for samples with $B/B_{\phi} = 1.0$, $F_{p} = 1.5$, and $F_{ac} = 1.5$,
  highlighting the enhanced effectiveness of the ratchet mechanism in
  the Conf array.
  (a) The trajectories for the positive half of the ac drive cycle in
  the RandG array showing the formation of disordered flow channels.
  (b) The negative half of the ac drive cycle in the RandG array has a
  similar pattern and density of flow channels. 
  (c) In the positive half of the ac drive cycle for the Conf array,
  the vortices cannot move past the densely pinned region.
  (d) In the negative half of the ac drive cycle in the Conf array,
  the vortices can easily funnel between the arches
  in the conformal array.
  (e) In the positive half of the ac drive cycle for the SquareG array,
  vortices can slip through the interstitial regions between pinned
  vortices.
  (f) Similar interstitial motion occurs in the negative
  half of the ac drive cycle for the SquareG array.
}
\label{fig:3}
\end{figure}

The relative effectiveness of the different arrays can be more 
clearly understood by plotting the trajectories of the vortices
during the positive and negative portions of a single ac cycle.  
Figure~\ref{fig:3}(a,b) shows the trajectories for both halves of the
ac cycle in a
a RandG array with $B/B_{\phi} = 1.0$, $F_{p} = 1.5$ and $F_{ac} = 1.5$.
Under both positive and negative drive,
the vortices form disordered flow paths with a similar density that is
independent of the driving direction.
In contrast, in Fig.~\ref{fig:3}(c) during the positive portion of the
ac driving cycle for the Conf array,
almost no vortices can cross the densely pinned regions of the sample;
instead, the vortices either become
trapped at pining sites or remain localized in interstitial cages formed
by the pinned vortices.
Figure~\ref{fig:3}(d) shows that in the negative portion of the cycle
for the Conf array,
numerous vortices move into the interstitial 
regions and funnel through the conformal arch structures, producing
significant vortex  motion in the negative $x$ direction.
In the SquareG system, Figs.~\ref{fig:3}(e,f) show that
vortex motion is strongly suppressed for both directions of drive
and occurs only when
interstitial vortices manage to squeeze between equally spaced
occupied pinning sites.  The barrier to this type of vortex motion is
the same in each half of the cycle.
In contrast, for the Conf
array the perpendicular spacing between pinned vortices in the sparse
portion of the array is larger than the equivalent spacing between pinned
vortices in the dense portion of the array,
so the interstitial
vortices experience much
different effective caging barriers
when entering the sparse side of the array than when
entering the dense side of the array.
In the RandG arrays, channels of 
easy vortex flow 
occur somewhere in the sample with equal probability
for both the positive and negative portions of the ac drive cycle.

\begin{figure}
\includegraphics[width=3.5in]{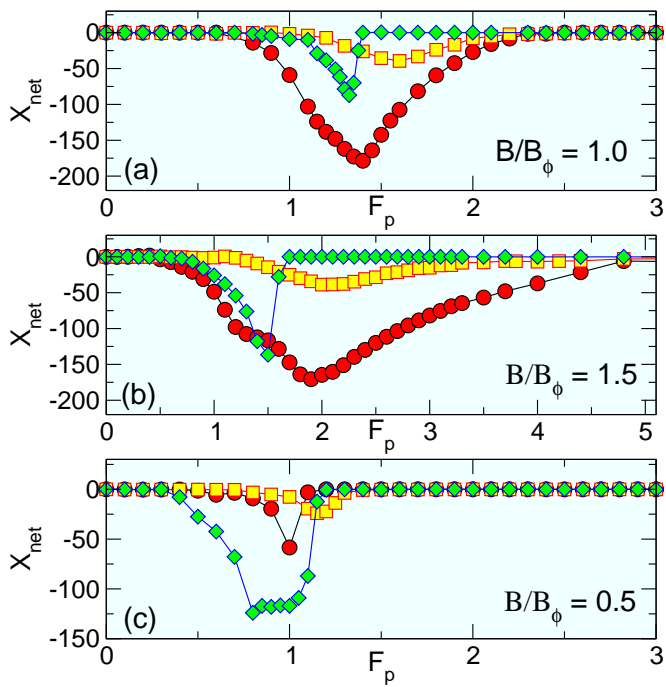}
\caption{
  $X_{\rm net}$, the average net displacement per vortex
  after 50 ac drive cycles, vs $F_{p}$ for the Conf (red circles),
  RandG (yellow squares), and SquareG (green diamonds) arrays.
  Here $F_{ac} = 0.7$ and $n_{p} = 1.0$.
  In general, the ratchet effect is suppressed for weak pinning and
  for strong pinning.
  (a) At $B/B_{\phi} = 1.0$, the Conf array exhibits the strongest ratchet
  effects, followed by the SquareG array.  The RandG array has the
  weakest ratchet effect.
  (b) At $B/B_{\phi} = 1.5$ the ratchet effect extends to higher
  values of $F_{p}$ in all the systems. The Conf ratchet is still
  the most effective.
  (c) At $B/B_{\phi} = 0.5$, the SquareG ratchet is more effective than the
  Conf or RandG ratchets.
}
\label{fig:4}
\end{figure}

In Fig.~\ref{fig:4}(a) we plot $X_{\rm net}$, the average net displacement
per vortex after 50 ac drive cycles, versus $F_{p}$ 
for Conf, RandG, and SquareG samples with $B/B_{\phi} = 1.0$ and $F_{ac} = 0.7$.  
For weak pinning $F_{p} < 0.7$, the vortices move elastically and easily
slide over the pinning sites so that there
is no ratchet effect in any of the arrays.
For $F_{p} > 1.5$ most of the vortices become increasingly pinned 
and the ratchet effect is reduced.
The optimal ratchet effect occurs for the Conf array
at $F_{p} = 1.4$, where 
there is a mixture of pinned vortices
coexisting with vortices that move temporarily through the
interstitial regions as illustrated
in Fig.~\ref{fig:3}(c,d).
The SquareG array has a weaker ratchet effect in the range
$0.8 < F_{p} < 1.4$, with a relatively sharp cutoff at the upper
end of this range that occurs when
the ability of the pinned vortices to shift inside the pinning sites
is reduced, preventing the
interstitial vortices from slipping between
occupied pinning and causing the
motion to become localized, as shown
in Fig.~\ref{fig:3}(e,f).
There is a weak ratchet effect for 
the RandG array
with an extremum at $F_{p} = 1.6$ where the combination of the ac
drive and the vortex-vortex interactions causes
a portion of the vortices to depin.
The maximum magnitude of the ratchet effect for the RandG array is
smaller than that for the SquareG array; however, the effect
occurs over a wider range of $F_{p}$.
Figure \ref{fig:4}(b) shows
$X_{\rm net}$ versus $F_p$ for $B/B_{\phi} = 1.5$, where there are
more interstitial vortices.
Here the range of $F_{p}$ 
over which the ratchet effect occurs for the Conf and RandG arrays
extends up to $F_p=5.0$, with the ratcheting
for $F_{p} > 2.1$ completely dominated by the flow of interstitial vortices.
For the SquareG array the ratchet effect is lost for
$F_{p} > 1.7$, the point at which the interstitial vortices can no longer
slip through the one-dimensional interstitial channels of the array.
In Fig.~\ref{fig:4}(c) we plot $X_{\rm net}$ versus $F_p$ at
$B/B_{\phi} = 0.5$, where there are few interstitial vortices.
Here most of the motion occurs when vortices jump
from one pinning site to another. 
The ratchet effect for all three arrays vanishes for $F_{p} > 1.4$ when
vortex hopping is suppressed.
At this vortex density, the
ratchet effect is most pronounced for the SquareG array,
where the vortices are able to hop along one-dimensional channels of
pinning sites.

\begin{figure}
\includegraphics[width=3.5in]{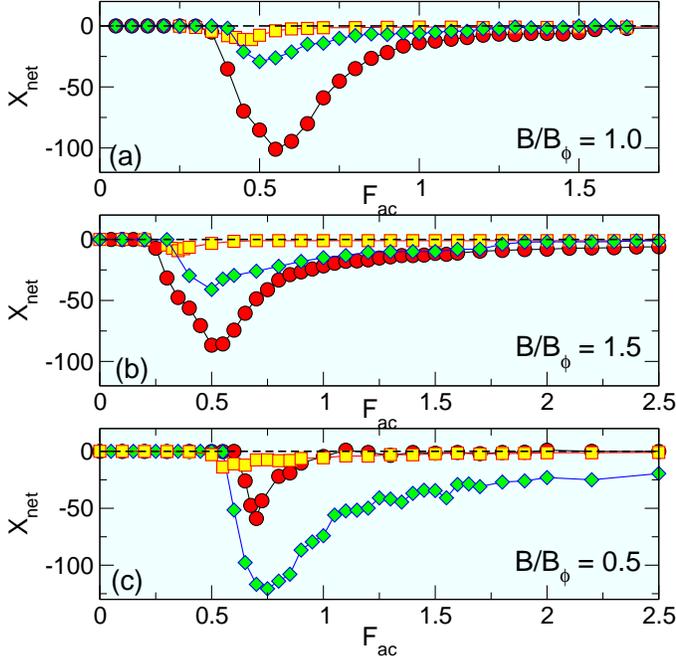}
\caption{
  $X_{\rm net}$ vs ac amplitude $F_{ac}$
  for Conf (red circles), RandG (yellow squares), and SquareG (green diamonds)
  arrays with $F_{p} = 1.0$ and $n_{p} = 1.0$.
  (a) At $B/B_{\phi} = 1.0$, the ratchet effect is reduced at low
  $F_{ac}$ when the vortices are pinned as well as at higher $F_{ac}$
  when the vortices move rapidly over the pinning array.
  (b) $B/B_{\phi} = 1.5$.
  (c) At $B/B_{\phi} = 0.5$ the SquareG array produces the
  most effective ratchet.
}
\label{fig:5}
\end{figure}

In Fig.~\ref{fig:5}(a) we plot $X_{\rm net}$ versus
the ac drive amplitude $F_{ac}$ for Conf, RandG, and SquareG
samples with $F_{p} = 1.0$ and $B/B_{\phi} = 1.0$.
For $F_{ac} < 0.35$, the vortices are mostly pinned and
the ratchet effect is absent for all three of the pinning geometries.
At intermediate $F_{ac}$ the Conf array has
the strongest ratchet effect, with an extremum
in $X_{\rm net}$ at $F_{ac}  = 0.55$. The SquareG array
has the next most effective ratchet effect, with an optimal
magnitude at $F_{ac} = 0.5$.
For higher values of $F_{ac}$, the
vortices are all in motion during some portion of the
driving cycle and the ratchet effect gradually decreases to zero
with increasing $F_{ac}$. We observe
a similar trend at $B/B_{\phi} = 1.5$ as shown in Fig.~\ref{fig:5}(b).
Here the ratchet effect for the Conf array
extends  
up to much larger values of $F_{ac}$;
however, the maximum value of $|X_{\rm net}|$
is slightly smaller than for the $B/B_{\phi} = 1.0$
case.
For $B/B_{\phi} = 0.5$ in Fig.~\ref{fig:5}(c),
the dominant motion is hopping of vortices from pinning site to pinning site.
Here the ratchet effect is strongest for the SquareG array,
similar to what is shown in Fig.~\ref{fig:4}(c).

\begin{figure}
\includegraphics[width=3.5in]{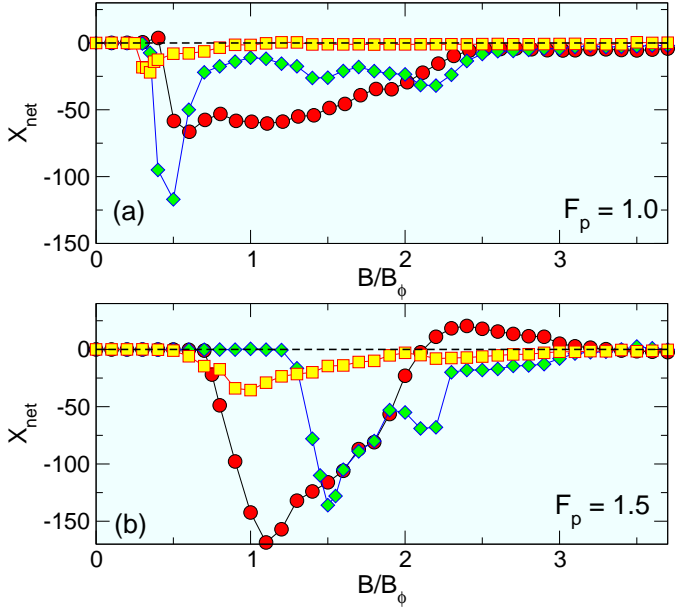}
\caption{
  $X_{\rm net}$ vs $B/B_{\phi}$ for Conf (red circles), RandG (yellow squares),
  and SquareG (green diamonds) arrays with
  $F_{ac} = 0.7$.
  (a) At $F_{p} = 1.0$ the ratchet effect is negative for the
  entire range of $B/B_{\phi}$.
  (b) At $F_{p} = 1.5$ there is a reversal in the ratchet effect
  for the Conf array for $2.125 < B/B_{\phi} < 3.375$.     
}
\label{fig:6}
\end{figure}

\section{Ratchet Reversal}

In Fig.~\ref{fig:6}(a) we plot $X_{\rm net}$ versus $B/B_{\phi}$
for Conf, RandG, and SquareG arrays with $F_{p} = 1.0$ and $F_{ac} = 0.7$.
Here the Conf array outperforms the RandG array for all fields and
the SquareG array for $0.6 < B/B_{\phi} < 2.0$.
For $B/B_{\phi} > 2.0$ the
vortex-vortex interactions become dominant
and the ratchet effect is suppressed in all the arrays.
In the SquareG array,
due to the periodic ordering along the $y$ direction,
some commensuration effects occur, 
such as enhanced pinning
near $B/B_{\phi} = 1.0$ which locally suppresses the ratchet effect.  

\begin{figure}
\includegraphics[width=3.5in]{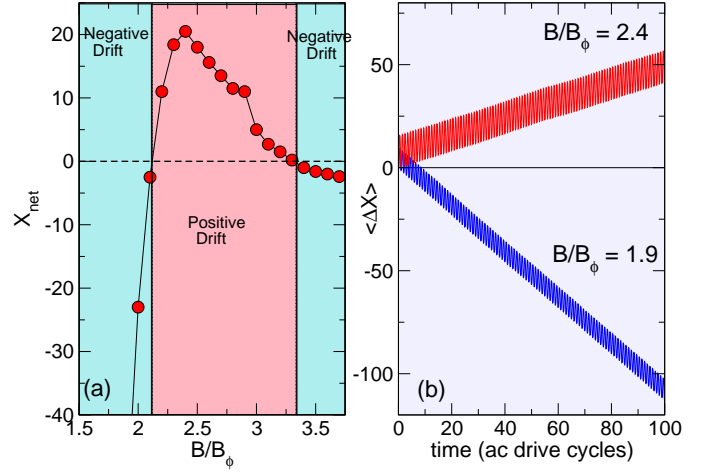}
\caption{
  (a) $X_{\rm net}$ vs $B/B_{\phi}$ for the Conf array in
  Fig.~\ref{fig:6}(b) highlighting the vortex ratchet reversal effect from 
  negative for $B/B_{\phi} < 2.25$ to positive
  for $2.25 <  B/B_{\phi} < 3.375$ to negative again at higher fields. 
  (b) $\langle \Delta X\rangle$ vs time in ac drive cycle numbers
  for the system in (a) at $B/B_{\phi} = 1.9$ (lower blue curve)
  where the vortex
  motion is in the negative $x$ direction and
  at $B/B_{\phi} = 2.4$ (upper red curve) where the motion is in the
  positive $x$ direction.  
}
\label{fig:7}
\end{figure}

Figure \ref{fig:6}(b) shows $X_{\rm net}$ versus $B/B_{\phi}$
for $F_{p} = 1.5$.  In this case, the ratchet effect for the SquareG
array is lost for $B/B_{\phi} < 1.2$ when the
vortices become strongly pinned at the pinning sites.
In general, the ratchet effect for the Conf array is stronger than that 
for the SquareG and RandG arrays, with a local extremum
for the ratchet effect in the negative or normal
direction occurring at $B/B_{\phi} = 1.05$.
The SquareG array has a local extremum in $X_{\rm net}$
in the negative direction at $B/B_{\phi} = 1.5$, followed by a sharp
drop in $X_{\rm net}$
for $B/B_{\phi} > 2.25$.
We find a ratchet reversal in the Conf array, where $X_{\rm net}$ switches
from negative to positive over the range
$2.1 < B/B_{\phi} < 3.375$.
There is a local maximum in the positive ratchet effect at $B/B_{\phi} = 2.4$.
In Fig.~\ref{fig:7}(a) we show a highlight of
$X_{\rm net}$ versus $B/B_{\phi}$ from Fig.~\ref{fig:6}(b) for
the Conf array indicating that two reversals in the ratchet effect occur.
Figure~\ref{fig:7}(b) illustrates $\langle \Delta X\rangle$ 
vs time in ac drive cycles for the system in Fig.~\ref{fig:7}(a) 
at $B/B_{\phi} = 1.9$, where the motion is in the negative
$x$ direction, and at $B/B_{\phi} = 2.4$, where the motion is in the positive
$x$ direction, showing 
more clearly the change in the net direction of vortex motion.      

\begin{figure}
\includegraphics[width=3.5in]{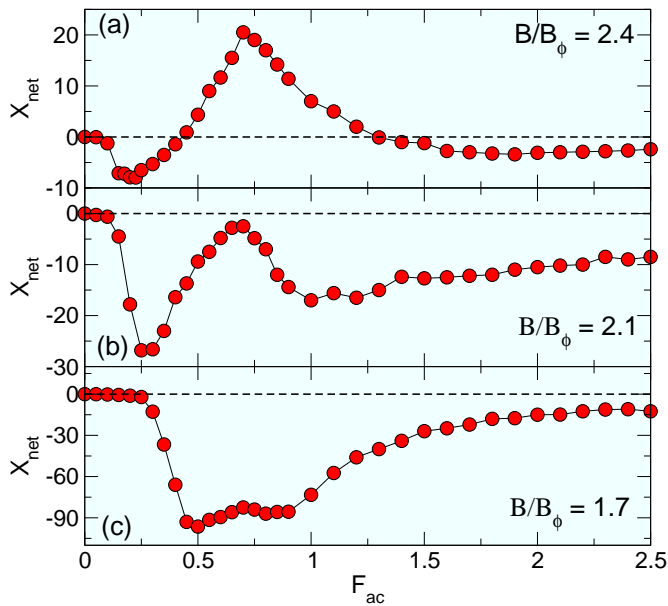}
\caption{$X_{\rm net}$ vs $F_{ac}$ for Conf arrays with $F_{p} = 1.5$.
  (a) At $B/B_{\phi} = 2.4$, there is a transition from a negative
  ratchet effect at low $F_{ac}$ to a positive ratchet effect,
  followed by a second transition
  back to a negative ratchet effect.
  (b) At $B/B_{\phi} = 2.1$ the ratchet effect is always negative;
  however, there is a local minimum and a local maximum of the ratchet effect.
  (c) At $B/B_{\phi} = 1.7$, the ratchet effect is always negative and
  has few features.
}
\label{fig:8}
\end{figure}

In Fig.~\ref{fig:8}(a) we plot
$X_{\rm net}$ versus $F_{ac}$ for a Conf array with
$F_{p} =  1.5$ at $B/B_{\phi} = 2.4$, where
there are  multiple reversals in the
ratchet effect.
For $F_{ac} < 0.1$ there is no ratchet effect since
the vortices move only small distances. 
A negative ratchet effect occurs for $0.1  < F_{ac} <  0.45$,
while for $0.45 \leq F_{ac} < 1.3$ there is a positive ratchet effect
with a maximum amplitude at $F_{ac} = 0.7$. 
There is another transition
to a weaker negative ratchet effect for $F_{ac} > 0.13$,
and $X_{\rm net}$ gradually approaches zero for high values of $F_{ac}$.
Figure~\ref{fig:8}(b) shows that at $B/B_{\phi} = 2.1$,
the ratchet effect is always negative; however, there
are still local features in the response such as at
$0.3 < F_{ac} < 1.0$ where the negative ratchet effect is strongly reduced.
In Fig.~\ref{fig:8}(c) at $B/B_{\phi} = 1.7$, the ratchet effect is
strongly negative with an extremum in $X_{\rm net}$
near $F_{ac}=0.5$.  The ratchet effect goes to zero for increasing $F_{ac}$.

\begin{figure}
\includegraphics[width=3.5in]{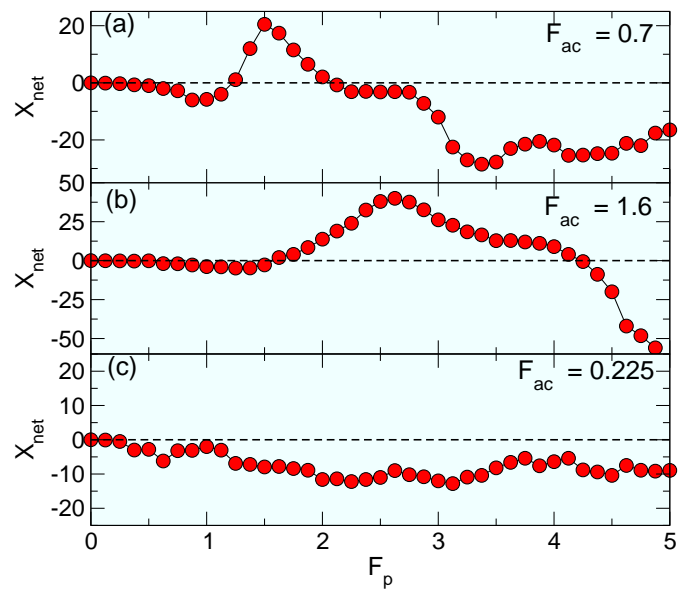}
\caption{
  $X_{\rm net}$ vs $F_{p}$ for Conf arrays with $B/B_{\phi} = 2.4$.
  (a) At $F_{ac} = 0.7$ there are multiple reversals as $F_{p}$ increases. (b)  
  At $F_{ac} = 1.6$ there are again multiple reversals and the positive
  ratchet effect extends over a wider range of $F_{p}$.
  (c) At $F_{ac} = 0.225$ there is a weak negative ratchet effect. 
}
\label{fig:9}
\end{figure}

In Fig.~\ref{fig:9}(a) we show $X_{\rm net}$ versus $F_{p}$ for a Conf array
at $B/B_{\phi} = 2.4$ and $F_{ac} = 0.7$.
There is a negative ratchet effect
for $0 <  F_{p} < 1.25$, a positive ratchet effect for
$1.25 \leq F_{p} < 2.1$, and a much larger negative ratchet effect for
$F_{p} > 3.0$.
At intermediate $F_p$ when there is a
positive ratchet effect, vortices can be temporarily trapped by
pinning sites.
The negative ratchet effect for large $F_p$ arises from the interstitial
flow of vortices, and $X_{\rm net}$ saturates at large $F_p$ since the
caging barrier experienced by interstitial vortices from the neighboring
pinned vortices does not increase with increasing $F_p$.
In Fig.~\ref{fig:9}(b), at $F_{ac} = 1.6$ there is a negative ratchet effect
for $0 < F_{p} < 1.625$, a positive ratchet effect for
$1.625 \leq F_{p} < 4.25$, and another negative ratchet regime
for $F_{p} > 4.25$. 
The vortices at the pinning sites remain permanently pinned for
$F_{p} > 4.25$.   
The positive ratchet effect is larger and extends out to higher values of
$F_p$ for the $F_{ac}=1.6$ system compared to the $F_{ac}=0.7$ system.
Figure~\ref{fig:9}(c) shows that at $F_{ac} = 0.225$,
there is a weak negative ratchet effect for all values of $F_{p}$.
Although we focus here on the Conf array, we also found that
some weak ratchet reversals are possible in the SquareG array; however,
we did not observe a vortex ratchet reversal for the RandG array.

\begin{figure}
\includegraphics[width=3.5in]{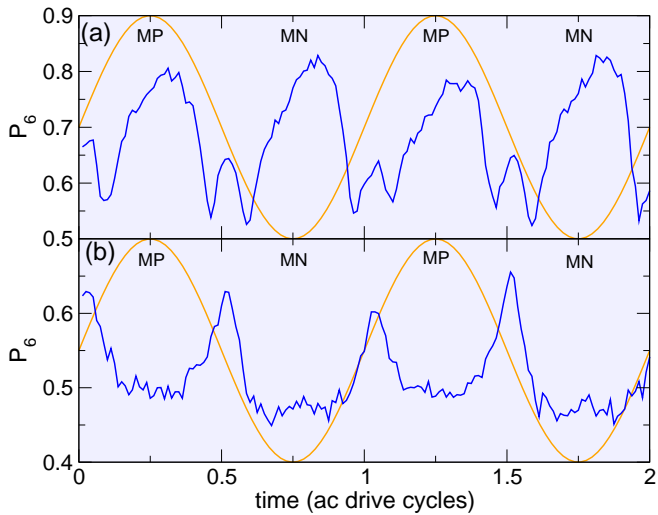}
\caption{
  Dark blue lines: $P_6$, the fraction of sixfold-coordinated particles,
  vs time in ac drive
  cycles for Conf arrays from Fig.~\ref{fig:9}(a) with $B/B_{\phi}=2.4$
  and $F_{ac}=0.7$.
  Light orange lines indicate the phase of the drive cycle.  MP is the maximum
  positive drive and MN is the maximum negative drive.
  (a) At $F_{p} = 0.875$ the ratchet effect is negative. The system is most 
  ordered whenever the magnitude of the ac drive is maximum; however, the
  ordering peaks for the negative portions of the drive cycle are slightly
  higher than those for the positive portions of the drive cycle.
  (b) At $F_{p} = 1.5$ the ratchet effect is positive.
  The system is most ordered whenever the magnitude of the ac drive is
  close to zero, but the net motion is determined by the relatively
  larger ordering at MP points compared to MN points.
}
\label{fig:10}
\end{figure}

In order to better understand the vortex dynamics and ordering
during an individual ac cycle,
in Fig.~\ref{fig:10} we plot the time series of the fraction of
sixfold-coordinated vortices,
$P_6=N_{v}^{-1}\sum_{i=1}^{N_v}\delta(z_i-6)$, versus time.  Here $z_i$, the
coordination number of vortex $i$, is obtained from a Voronoi construction.
Superimposed over the plot is a curve showing the phase of the
ac drive, and the points at which the drive reaches its maximum positive
value are marked MP while those at which the drive reaches its maximum
negative value are marked MN.
Figure~\ref{fig:10}(a) shows $P_{6}$ versus time for
the system from Fig.~\ref{fig:9}(a) with $B/B_{\phi} = 2.4$ and $F_{ac} =0.7$
at $F_{p} = 0.875$ where there is a negative ratchet effect.
Here $P_{6} = 0.65$
at the start of each drive cycle when the drive magnitude is zero,
deceases slightly when the drive becomes positive and the system
disorders, then reaches its highest values of $P_{6} \approx 0.8$ 
in the MP portions of the drive cycle and $P_6\approx 0.83$ in the
MN portions of the drive cycle.
When the magnitude of the ac drive is maximum,
all the vortices move
elastically, and since they are slightly more ordered during the
negative cycle of the drive,
they can slide slightly further in the negative $x$ direction than in the
positive $x$ direction, giving a negative ratchet effect.
At $F_{p} = 1.5$ in
Fig.~\ref{fig:10}(b), the ratchet effect is positive 
and the vortex ordering is reversed.
The vortices are now the most ordered when the magnitude of the ac drive is
close to zero, and they are disordered when the ac drive magnitude reaches
a maximum.
During the MP portion of the drive cycle, 
$P_{6} = 0.5$, while in the MN portion of the drive cycle
the system is more disordered with $P_{6} = 0.47$.
The more ordered vortices are able to slide slightly further in the
positive $x$ direction, resulting in a net positive ratchet effect.
There is also an asymmetry in the ordering at the zero force portions
of the drive cycle.
The value of $P_6$ at cycle times of $0.0$ and $1.0$ is smaller than
that at times of $0.5$ and $1.5$; however, since the vortices are not
moving during this portion
of the cycle, this asymmetry does not produce a preferred direction of motion.
In general we find that the ordering of the vortices at the MP and MN points
of the drive determines the direction of the ratchet motion, with the
net ratchet effect occurring in whichever drive direction generates the
most ordered vortex arrangement.

\section{Thermal Effects}

\begin{figure}
\includegraphics[width=3.5in]{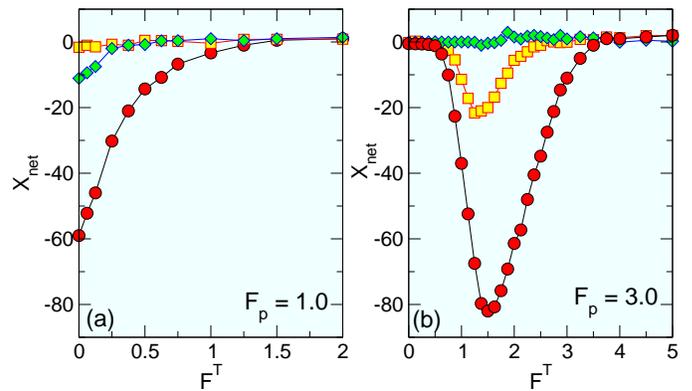}
\caption{
  $X_{\rm net}$ vs $F^{T}$ for Conf (red circles), RandG (yellow squares),
  and SquareG (green diamonds) arrays with $B/B_{\phi} = 1.0$ and $F_{ac} = 0.7$.
  (a) At $F_{p} = 1.0$, thermal effects reduce the ratchet effect.
  (b) At $F_{p} = 3.0$ thermal effects can
increase the ratchet effect over a range of $F^{T}$.  
}
\label{fig:11}
\end{figure}

We next consider thermal effects on the ratchet response.
For weak pinning, the addition of thermal fluctuations
monotonically decreases the ratchet effect for all three geometries,
as shown in Fig.~\ref{fig:11}(a) for $F_{p} = 1.0$, $F_{ac} = 0.7$,
and $B/B_{\phi} = 1.0$.
As before, the ratchet effect is most pronounced
for the conformal array.
As $F_{p}$ increases, the vortices become more strongly pinned, and
the addition of thermal fluctuations can increase the ratchet effect by
permitting vortices to escape from pinning sites or
interstitial caging sites via thermal activation.
In Fig.~\ref{fig:11}(b) we plot $X_{\rm net}$ versus $F^{T}$ in samples
with $F_p$ increased to $F_p=3.0$, showing a strong ratchet effect in
the Conf array.
Here, the ratchet effect is lost at small $F^{T}$ since the 
vortices are strongly pinned, and the ratchet effect also disappears
for high values of $F^{T}$ when the thermal fluctuations become so strong that 
the vortices enter a molten state that interacts too weakly with the
substrate for an asymmetry in the response to positive and negative drives
to be noticeable.
The largest ratchet signatures appear for
intermediate $F^{T}$. In general, when $F_{p}$ increases, the
point at which the magnitude of the ratchet effect is largest shifts to
higher values of $F^T$.

\begin{figure}
\includegraphics[width=3.5in]{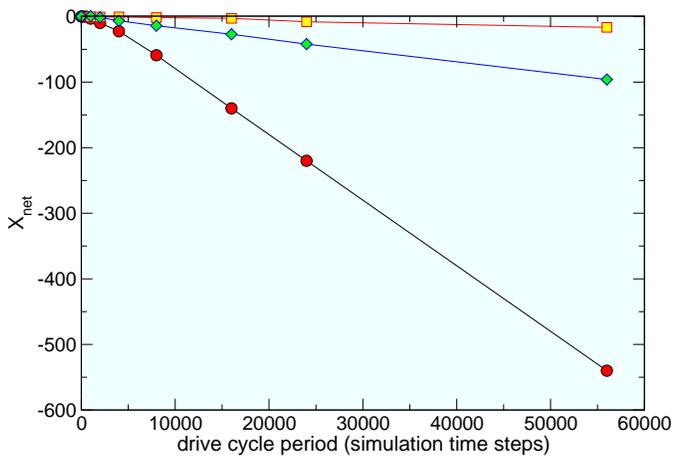}
\caption{
$X_{\rm net}$ versus ac cycle period in simulation time steps
  for Conf (red circles), RandG (yellow squares), and SquareG (green diamonds)
  arrays with $B/B_{\phi} = 1.0$, $F_{ac} = 0.7$, and $F_{p} = 1.0$.
$X_{\rm net}$ increases linearly with the drive cycle period.
}
\label{fig:12}
\end{figure}

We find that $X_{\rm net}$ increases linearly as the period of the ac driving
cycle increases, as illustrated in
Fig.~\ref{fig:12} for Conf, RandG, and SquareG arrays with
$B/B_{\phi} = 1.0$, $F_{p} = 1.0$, and $F_{ac} = 0.7$.
As the other parameters are varied, we find a robust increase in the
magnitude of the ratchet effect with decreasing ac frequency.

\section{Ratchet Effects for Colloidal Particles}

\begin{figure}
\includegraphics[width=3.5in]{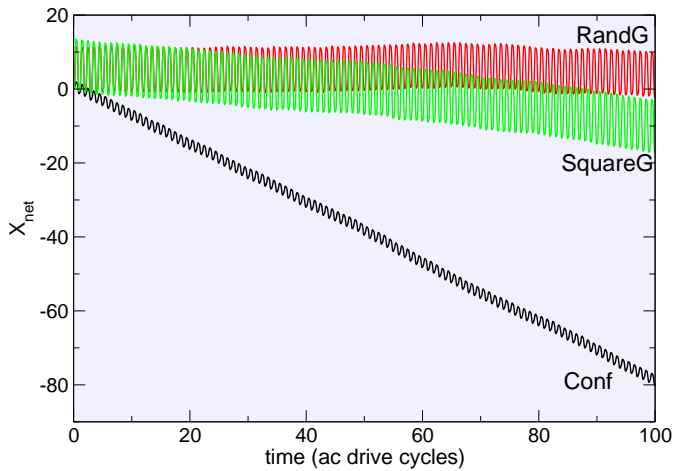}
\caption{
  $\langle \Delta X\rangle$ vs time measured in ac drive cycles
  for colloidal particles interacting with a Conf array (bottom black curve),
  a SquareG array (middle green curve), and a RandG array (upper red curve)
  with $N_{c}/N_{p} = 1.0$, $F_{p} = 1.0$, $F_{ac} = 0.55$, and
  $A_{c} = 0.01$.  As in the vortex case shown in Fig.~\ref{fig:2}, the Conf
array produces the strongest ratchet effect.  
}
\label{fig:13}
\end{figure}

\begin{figure}
\includegraphics[width=3.5in]{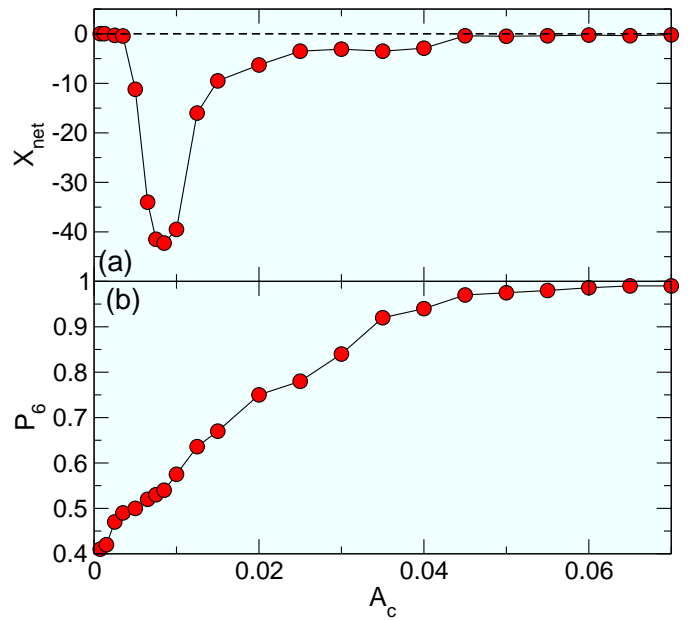}
\caption{
  (a) $X_{\rm net}$ vs colloid-colloid interaction strength
  $A_{c}$ for the Conf colloid system in Fig.~\ref{fig:13}
  with $N_{c}/N_{p} = 1.0$, $F_{p} = 1.0$, and $F_{ac} = 0.55$.
  (b) The corresponding fraction of sixfold coordinated colloids
  $P_{6}$ vs $A_{c}$.
  Here the maximum ratchet effect occurs when $P_{6} = 0.6$,
  indicating that although colloid-colloid interactions remain
  important, the system is in a disordered state. When the
  system forms a crystalline state with $P_{6} \approx 1.0$,
  the ratchet effect disappears.
}
\label{fig:14}
\end{figure}

Ratchet effects can be generated in systems of
colloidal particles interacting with various types
of periodic arrays of traps that are created using
optical means \cite{46,47,48,49}.
The ability to make structures similar to conformal lattices has been
demonstrated by Xiao {\it et al.} \cite{50}, who examined a
colloidal ratchet effect on optical traps forming Fibonacci spirals.
In that case the ratchet effect is induced by rotating the
potential through a three-step cycle.
Here we consider the ac-driven motion of colloidal particles over a
Conf array.
The equation of motion for colloids is similar to that given
in Eqn.~1 for vortices, except that
the pairwise repulsive colloid-colloid interaction
potential has the form $V(R_{ij}) = A_{c}\exp(-\kappa R_{ij})/R_{ij}$, 
where $E_{0} = q^2Z^{*2}/4\pi\epsilon\epsilon_{0}a_{0}$,
$q$ is the dimensionless interaction strength, 
$Z^{*}$ is the effective charge of the colloidal particles,
$\epsilon$ is the solvent dielectric constant, 
and $1/\kappa$ is the screening length which we set equal to $1.0$.
The number of colloids in the sample is $N_c$.
For our parameters, the interactions between colloids for $R_{ij}<1$ is
much larger than the interactions between vortices separated by the
same distance, 
while for $R_{ij}>1$ the colloidal interaction strength
falls off much more rapidly than the vortex-vortex interaction strength,
so that nearest neighbor interactions are dominant in the colloidal system.
We have conduced a series of simulations
for colloidal particles moving through Conf, RandG, and SquareG
pinning landscapes under an ac driving force, and find results very similar
to those obtained in the vortex system.
For example,
in Fig.~\ref{fig:13} we plot $\langle \Delta X \rangle$ versus time for
colloids interacting with Conf, SquareG, and RandG arrays
for $N_{c}/N_{p} = 1.0$, $F_{p}=1.0$, $F_{ac} = 0.55$, and $A_{c} = 0.01$,
where we observe that just as in the vortex case, the Conf array
produces the most pronounced ratchet effect, the SquareG array shows
a weak ratchet effect, and the RandG array does not exhibit a ratchet effect.
Since the effective charge on the colloids
can be changed experimentally, it is possible
to hold the substrate strength fixed and modify how strongly the 
colloids interact with one another.
In Fig.~\ref{fig:14}(a) we plot $X_{\rm net}$ versus $A_{c}$ for the
Conf array 
from Fig.~\ref{fig:13}, and in Fig.~\ref{fig:14}(b) we show
the corresponding fraction of sixfold coordinated colloids $P_{6}$
versus $A_{c}$. 
For $A_{c} = 0$ the colloids all become pinned
in the pinning sites since $F_{ac} < F_{p}$.   
As $A_{c}$ increases, the colloid-colloid interactions
become important and a
ratchet effect arises
with a maximum amplitude near
$A_{c} = 1.0$.
The largest ratchet effect is associated with a sixfold ordering fraction of
$P_{6} = 0.6$, indicating that the 
colloids are still disordered, with some colloids trapped in pinning sites
and others occupying interstitial regions between pins.
For $A_{c} > 0.01$ the ratchet effect begins to diminish with increasing
$A_c$ while simultaneously
$P_{6}$ increases, indicating an increase in the ordering
of the colloids.
For $A_{c} > 0.04$, the colloids form a rigid triangular
lattice as indicated by the fact that $P_{6} \approx 1.0$,
and the ratchet effect disappears.
These results show that in order for a ratchet effect to appear
in the gradient pinning arrangements, it is generally necessary for
plasticity or defects in the colloid or vortex lattice to appear.
Our results with the colloidal system indicate that pronounced ratchet
effects should be realizable in a variety
of systems where assemblies of interacting particles are driven with an
ac drive over conformal array substrates.

\section{Summary}

We examine ratchet effects for ac driven vortices interacting with
a conformal pinning array and with square and random pinning arrays containing a
gradient along one direction.
In general, the conformal pinning array produces the most pronounced
ratchet effect, particularly for fields greater than the first matching
field.
The enhanced effectiveness of the conformal ratchet results in part from
the fact that the pinning sites in the low density portion of the array
are widely spaced not only parallel to but also perpendicular to the net
pinning gradient direction, permitting the easy flow of interstitial vortices
through the sparse portion of the array.
In contrast, for the square pinning array with a gradient,
the perpendicular distance between pinning sites is constant throughout
the array, producing the same barrier for interstitial motion in both the
sparse and dense portions of the array and reducing the relative magnitude
of the
ratchet effect for fields at which interstitial vortices are present.
For the random pinning array with a gradient, channels of easy vortex flow
form for driving in either direction, significantly reducing the effective
asymmetry of the array.
We find that the
conformal array exhibits a series of vortex ratchet reversals
as a function of vortex density, ac drive amplitude,
and pinning strength, and we show that the direction of the ratchet is
determined by the amount of order present in the vortex lattice at
different phases of the ac driving cycle.
Finally, we demonstrate that the conformal array also produces a larger
ratchet effect compared to square and random pinning arrays with a gradient
in systems of colloidal particles,
suggesting that pronounced
ratchet effects should be a general feature of
particles moving over conformal arrays.

\acknowledgments
This work was carried out under the auspices of the 
NNSA of the 
U.S. DoE
at 
LANL
under Contract No.
DE-AC52-06NA25396.

\end{document}